\documentclass[fleqn, usenatbib]{mnras}

\usepackage{newtxtext}
\usepackage{newtxmath}

\usepackage[T1]{fontenc}
\usepackage{pdflscape}
\usepackage{longtable}
\usepackage{booktabs}
\usepackage{color}
\usepackage{amssymb}
\usepackage{mathtools}
\usepackage{xspace}
\usepackage{rotating}
\usepackage{appendix}
\usepackage{multirow}
\usepackage{multicol}
\usepackage{array}
\usepackage{subcaption}
\usepackage{comment}
\usepackage{tabularx}
\usepackage{bigstrut}
\usepackage{threeparttable}
\usepackage{afterpage}
\usepackage{capt-of}

\usepackage[normalem]{ulem}

\newcommand{\src}{Cen X-3}

\DeclareRobustCommand{\VAN}[3]{#2}
\let\VANthebibliography\thebibliography
\def\thebibliography{\DeclareRobustCommand{\VAN}[3]{##3}\VANthebibliography}

\usepackage{graphicx}	

\newcommand{\suzaku}{\emph{Suzaku}}
\newcommand{\swift}{\emph{Swift}}
\newcommand{\nustar}{\emph{NuSTAR}}
\newcommand{\ginga}{\emph{GINGA-LAC}}
\newcommand{\beppo}{\emph{BeppoSAX}}
\newcommand{\rxte}{\emph{RXTE}}
\newcommand{\ecyc}[1]{\ensuremath{E_{\rm{C}}}}

\newcommand{\crsfeq}[1]{\ensuremath{
   \ecyc{#1}= 11.6\frac{B}{10^{12}\, \rm{G}}(1 + z)^{-1}
}}

\title[Cyclotron line in Cen X-3]{New measurements of the cyclotron line energy in Cen X-3}

\author[Tomar et al.]{
Gunjan Tomar, $^{1}$\thanks{E-mail: gunjan@rri.res.in}
Pragati Pradhan, $^{2}$
and Biswajit Paul$^{1}$
\\
$^{1}$Raman Research Institute, Astronomy and Astrophysics, C. V. Raman Avenue, Bangalore 560080. Karnataka, India\\
$^{2}${Massachusetts Institute of Technology, 77 Massachusetts Ave. Cambridge, MA 02139, USA}
}

\date{Accepted XXX. Received YYY; in original form ZZZ}

\pubyear{2020}

\begin{document}
\label{firstpage}
\pagerange{\pageref{firstpage}--\pageref{lastpage}}
\maketitle

\begin{abstract}
We report results from the analysis of data from two observations of the accreting binary X-ray pulsar \src~carried out with the broadband X-ray observatories \suzaku~and~\nustar. The pulse profile is dominated by a broad single peak and show some energy dependence with two additional weak pulse peaks at energies below 15 and 25\, keV respectively. The broadband X-ray spectrum for 0.8-60.0\, keV for \suzaku~ and 3.0-60.0 keV for \nustar~is fitted well with high energy cut-off power law model along with soft-excess, multiple iron emission lines and a cyclotron absorption. The cyclotron line energy is found to be $30.29^{+0.68}_{-0.61}$ keV and $29.22^{+0.28}_{-0.27}$ keV respectively in the \suzaku~ and \nustar~ spectra. We make a comparison of these two measurements with four previous measurements of CRSF in \src~ obtained with \emph{Ginga}, \beppo~ and \rxte. We find no evidence for a dependence of the CRSF on luminosity.
Except for one CRSF measurement with \beppo~, the remaining measurements are consistent with a CRSF energy in the range of 29.5 - 30.0\, keV over a luminosity range of 1.1-5.4 $\times 10^{37}$ ergs s$^{-1}$ different from several other sources that show considerable CRSF variation in the same luminosity range.
\end{abstract}




\section{Introduction}
X-ray binaries are the brightest X-ray sources in our galaxy. High mass X-ray binaries (HMXBs) are the binary star systems consisting of a neutron star or a black hole and a donor star with mass $\geq$ 10M$_\odot$. The persistent X-ray binaries are mostly powered by capture of matter from strong stellar wind of the companion star. For accretion onto a high magnetic field neutron star, the accreted material is channeled along the magnetic field lines onto the magnetic poles at relatively large distance from the surface of the compact object creating accretion funnels, where the X-ray emission is originated due to conversion of the gravitational potential energy of the infalling material to kinetic energy which is then released as X-rays due to shocks and dissipations into the accretion column and on the hot spots (\citealt{Basko}).
\\
Cyclotron resonant scattering features (CRSF) or simply, cyclotron lines are usually seen in absorption against the continuum spectrum in many of these sources. In such strong magnetic fields, the energies of the electrons are quantized into equi-spaced Landau levels. Resonant scattering of hard X-ray photons on such electrons in line-forming region, results in cyclotron lines.  CRSFs allow a direct measurement of the magnetic field since the energy of the former is directly related to the magnetic field as \crsfeq ~n keV, where B$_{12}$ is the magnetic field strength in units of 10$^{12}$ G, n is the Landau level number, and z is the gravitational redshift at the line-forming region. n=1 and n=2, 3, 4.. correspond to the fundamental and harmonic cyclotron lines, respectively.
\\
For a given source, the cyclotron line energy is often found to depend on X-ray luminosity L$_X$. The correlation between the two was first observed in the transient source V 0332+53, in which the line energy decreases with increasing X-ray flux (\citealt{Tsy}). A positive correlation of CRSF energy with luminosity was first discovered in Her X-1 by \citet{staubert2007}. Since then, several other X-ray pulsars have been studied for the CRSF energy-L$_X$ relationship and it has been established that CRSF energy-luminosity relation is bimodal with source luminosity (\citealt{Becker}) with the majority of sources showing a positive correlation (\citealt{staubert2019}). Although, an unambiguous explanation is still lacking, different theoretical models have been proposed to explain these dependencies for e.g the shock-height model \citet{Becker}, or  the reflection model of \citet{poutanen}.
\\
In this paper, we focus on one of the most luminous persistent accreting X-ray pulsars in our galaxy, Cen X-3, first discovered with a rocket-borne detector (\citealt{Chodil}). Extensive follow-up observations with Uhuru revealed the binary and pulsar nature of the system (\citealt{Giacconi}, \citealt{Schreier}). The system is an HMXB consisting of a neutron star of mass 1.21$\pm  $0.21 M$_\odot$ which emits pulsed X-rays every 4.8 s while orbiting an O-type supergiant v779 Cen (\citealt{Krzeminski}) of mass and radius of 20.5$\pm$0.7 M$_\odot$ and 12 R$_\odot$, respectively every 2.1 days. Although a strong stellar wind emanates from the supergiant, an accretion disk fed by Roche lobe overflow was inferred from optical light curves by \citet{tjemkes} in 1986, which supply most of the accretion fuel. The high luminosity of the star (~5.0 $\times 10^{37} $erg s$^{-1}$; \citealt{Suchy_2008}) also suggests that the predominant mode of accretion is via a disk, as does the detection of QPO from the source (\citealt{Takeshima}, \citealt{Raichur_08}). The distance to the binary system is estimated to be $\sim$ 8 kpc (\citealt{Krzeminski}).
\\
The pulse phase averaged spectrum of X-ray pulsars is generally characterized by a flat power law with an energy index of 0.5-1.5 upto a high-energy cutoff at 10.0-20.0 keV. The analysis of  BeppoSAX data shows that the broadband ( 0.12 - 100 keV ) spectrum of \src ~ is well fitted by an absorbed power-law continuum (photon index $\sim$ 1.2)  modified by a high-energy rollover (with cutoff energy $\sim$ 14 keV and e-folding energy $\sim$ 8 keV) smoothed around the cutoff energy with a soft excess below 1 keV interpreted as a blackbody with kT $\sim$ 0.1 keV \citep{Burderi}.  Additionally, iron  K-emission lines at 6.4 keV, 6.67 keV and 6.97 keV are also clearly resolved (\citealt{Ebiwasa}, \citealt{naik12}). The 6.4 keV emission line is identified with the fluorescent iron K$\alpha$ line due to reprocessing by cold circumstellar matter surrounding the binary system (\citealt{Nagase_89}), whereas 6.67 keV and 6.97 keV lines are considered to be originated in the highly photoionized accretion disk corona(\citealt{kallman_89}). The analysis of Ginga data \citep{Nagase} showed that the spectrum at high energy is better fitted with a Lorentzian shape high energy turn over which resembles the effects of cyclotron scattering feature, the presence of CRSF at $\sim$ 28 keV has been confirmed by \citet{Santangelo} which implies a magnetic field at the surface of the neutron star of B = 2.98 $\times$ 10$^{12}$ G taking into account the effects of the gravitational field close to it.
\\
In spite of being one of the brightest X-ray sources in the galaxy and one of the most studied accretion powered X-ray pulsars, the cyclotron line feature in Cen X-3, and especially its luminosity dependence is not studied extensively. Cen X-3 shows luminosity variation by a factor of upto 40 though it is a persistent X-ray source. Here we present timing and spectral analysis of two observations of Cen X-3 with the broadband X-ray missions Suzaku and NuStar. Along with detection of the cyclotron line in these two observations, we also compared the line energies with the previous reports, covering an X-ray luminosity range of 1.1-5.4 $\times 10^{37}$ ergs s$^{-1}$.

\section{Observations, Data Reduction and analysis}
\src~was observed once each  with the \emph{Suzaku} observatory \citet{M07} and the \nustar~(\emph{The Nuclear Spectroscopic Telescope Array}; \citealt{Harrison_2013}). The observation log is given in Table \ref{obsid}. An orbital phase resolved spectroscopy was carried out using the same \suzaku~ observation that reported many pre-eclipse dips (\citealt{Naik_2011}).

\begin{table}
\centering
\caption{Observation log for  
\suzaku~and \nustar}
\begin{tabular}{l l l l l}
\hline
ObsID & Start Date & Exposure & MJD\\
\hline
\hline
\multirow{2}{6em}{Suzaku 403046010} & & & & \\
& 2008-12-08 & 33.43 ks & 54808.37\\
\hline
\multirow{2}{6em}{NuSTAR 30101055002} & & & &\\
 & 2015-11-30 & 21.42 ks & 57356.75\\
\hline
\end{tabular}
\label{obsid}
\end{table}

\nustar~(\citealt{Harrison_2013}), launched in 2012 is the first hard X-ray focusing telescope operating in the energy band 3.0-79.0 keV. It consists of two co-aligned telescopes that focus X-ray photons onto their respective Focal Plane Modules (FPMA and FPMB). Each module consists of a two-by-two array of CdZnTe pixel detectors, each with 32$\times$32 0.6mm pixels which provide a 12' field of view, an angular resolution of 18" (FWHM) and energy resolution of 400 eV at 10 keV and 900 eV at 60 keV (FWHM). 
\\
The \nustar~ data was processed using \texttt{NUSTARDAS} (v1.8.0) along with \texttt{CALDB} v20190627. To extract the data products we used source-centered circular regions of 110" radius for both FPMA and FPMB. Background regions were extracted with an equivalent area away from the source. No contamination from stray light was observed. We then applied barycentric correction to the \nustar~ clean science event files to correct for X-ray photon arrival times using \texttt{FTOOL} \texttt{barycorr}. Light curves were extracted with a bin size of 20 ms and spectra from the corrected event files using the task \texttt{nuproducts}. The timing and spectral analysis was performed with \texttt{HEASOFT} v6.26.1 package.
\\
\emph{Suzaku} consists of two main payloads: the X-ray Imaging Spectrometer (XIS, 0.2-12 keV; \citealt{K07}) 
and the Hard X-ray Detector (HXD, 10-600 keV; \citealt{T07}). The XIS consists of four CCD detectors of which three (XIS 0, 2 and 3) are front illuminated 
(FI) and one (XIS 1) is back illuminated (BI). The HXD comprises PIN diodes and GSO crystal scintillator detectors. \\
The { \emph{Suzaku}   observation was made in 1/4 window mode} and data reduction was done on the filtered `cleaned' event files\footnote{http://heasarc.gsfc.nasa.gov/docs/suzaku/analysis/abc/} following the reduction technique mentioned in the Suzaku ABC guide. We applied barycentric correction to all event files using \texttt{aepipeline}. 
In case of CCD data as obtained by XIS, we had to take care of pile-up which is two photons of lower energy being read as one with higher energy which
may cause artifical hardening of the X-ray spectrum. {  
We discarded photons collected within the portion of the PSF where the estimated pile-up fraction was greater than 4 \% (determined using the FTOOLS task \texttt{pileest}) by choosing a source region with an annulus of $3^{'}$ radius and excluding the inner 15 arc-seconds of this annulus. Background for the XIS were extracted by selecting circular regions of the same size of $3^{'}$ radius  in a portion of the CCD that was not significantly contaminated by the 
source X-ray mission. XIS lightcurves and spectra were therefore extracted for these source and background regions. }

Being a photon counting detector, data from PIN detector have to be corrected for deadtime which is the time interval for which the detector 
electronics are processing one photon and thus cannot yet detect the arrival of another. This dead time correction was done using \texttt{FTOOLS} task \texttt{hxddtcor}. 
For the HXD/PIN, simulated `tuned' non X-ray background event files (NXB) corresponding to the month and year of the respective observations 
were used to estimate the non X-ray background \footnote{http://heasarc.nasa.gov/docs/suzaku/analysis/pinbgd.html}\citep{F09}. \\
The XIS spectra were extracted with 2048 channels and the PIN spectra with 255 channels.
Response files for the XIS were created using the CALDB version `20150312'. For the HXD/PIN spectrum, response files corresponding to the epoch 
of the observation were obtained from the \emph{Suzaku} guest observer
facility\footnote{http://heasarc.nasa.gov/docs/heasarc/caldb/suzaku/}. 

\subsection{Timing analysis}
\label{timing analysis}

\begin{figure}
\centering
\includegraphics[scale=0.4]{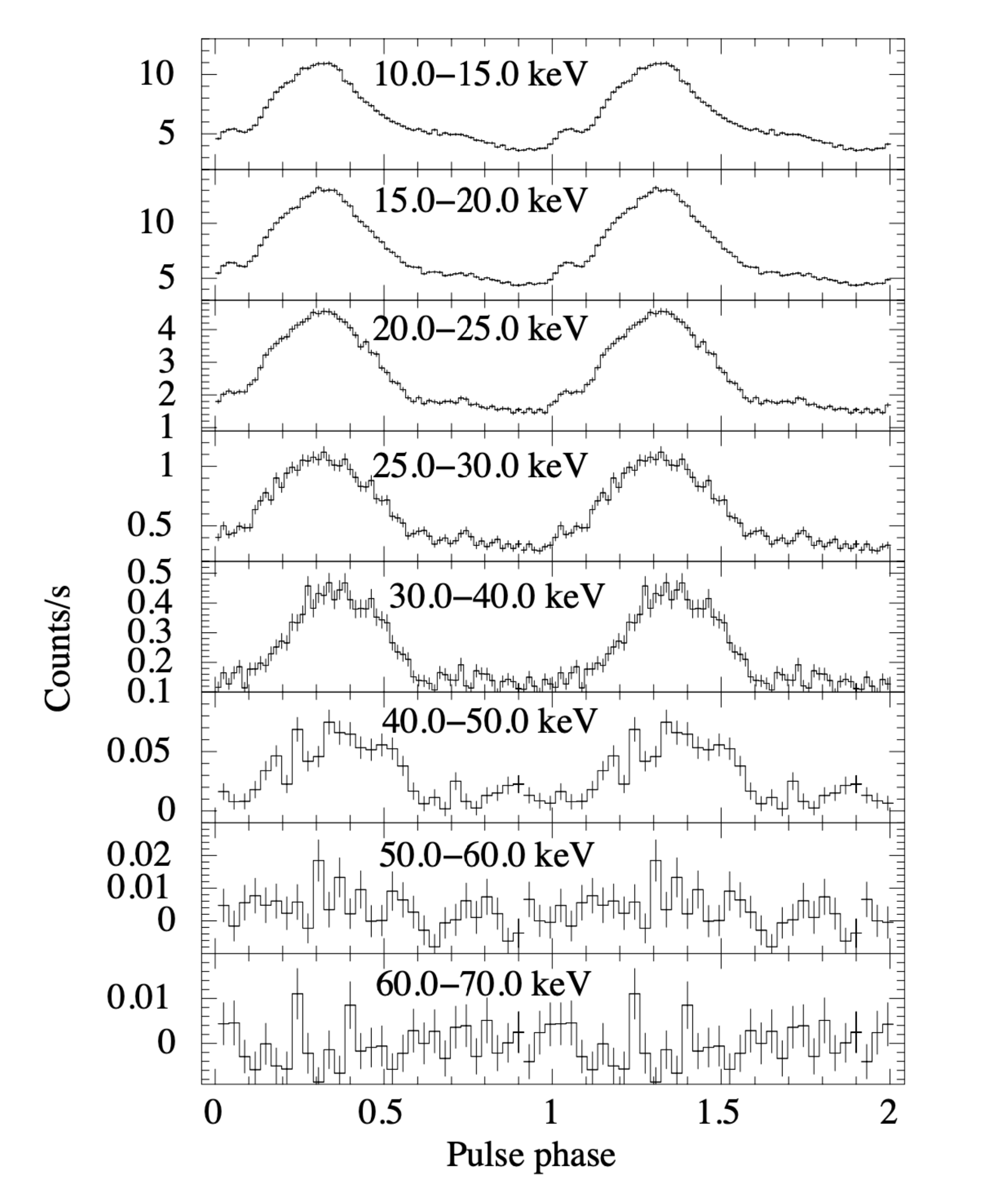}
\caption{Energy resolved pulse profiles generated for \src~ from \suzaku~HXD/PIN lightcurves. The lightcurves were corrected for orbital motion of the neutron star and barycenter corrected as well. See text for details. The corresponding energy range for each profile is noted in the inset.}
\label{pp_energy_su}
\end{figure}

\subsubsection{\suzaku~}
The \suzaku~XIS observation was carried out in 1/4 window mode which reads a part of the CCD array and has a time resolution of 2\, s. Since the spin period of \src~is $\sim$ 4.8\, s comparable to the time-resolution in this window mode, we are unable to use the XIS lightcurves for timing analysis and therefore only focus on the HXD/PIN lightcurves. The \suzaku~observation covered one full binary orbit of Cen X-3, including the eclipse and many absorption dips \citep{Naik_2011}. In the present work, data excluding the eclipse and dips have been considered (Segments D-H in Figure 3 of \citealt{Naik_2011})


We also corrected for orbital motion of the pulsar, by correcting for the arrival times in the PIN lightcurves using the orbital parameters from \citet{falagna2015} extrapolated to the observation time. Finally, the PIN lightcurves were background subtracted by generating a background lightcurve using the simulated background files
\footnote{http://heasarc.nasa.gov/docs/suzaku/analysis/pinbgd.html}\citep{F09}. 
 We use \texttt{efsearch} to determine a pulse period of the pulsar. In order to measure the period derivative, we first divided the light curve into smaller segments and noted the period in each segment and the corresponding lightcurve duration. The difference in the periods divided by the total duration of each light curve give an approximate value of period derivative. We further refined the period derivative by carrying out the pulse period determination repeatedly with different trial period derivatives to obtain maximum $\chi^{2}$ for a period derivative of -3.37$\times$10$^{-11}$ ss$^{-1}$ for spin period of 4.80461\, s ($\pm$ 0.00006) at MJD 54808. This value was used to fold the HXD lightcurves and obtain the energy-resolved pulse-profiles shown in Fig.~ \ref{pp_energy_su}. As seen in the figure, the pulse profiles are dominated by a single peak and show mild evolution with energy, especially the bump at $\phi$ = 0.05 below 25\, keV. 

\begin{figure}
\centering
\includegraphics[width=0.5\textwidth]{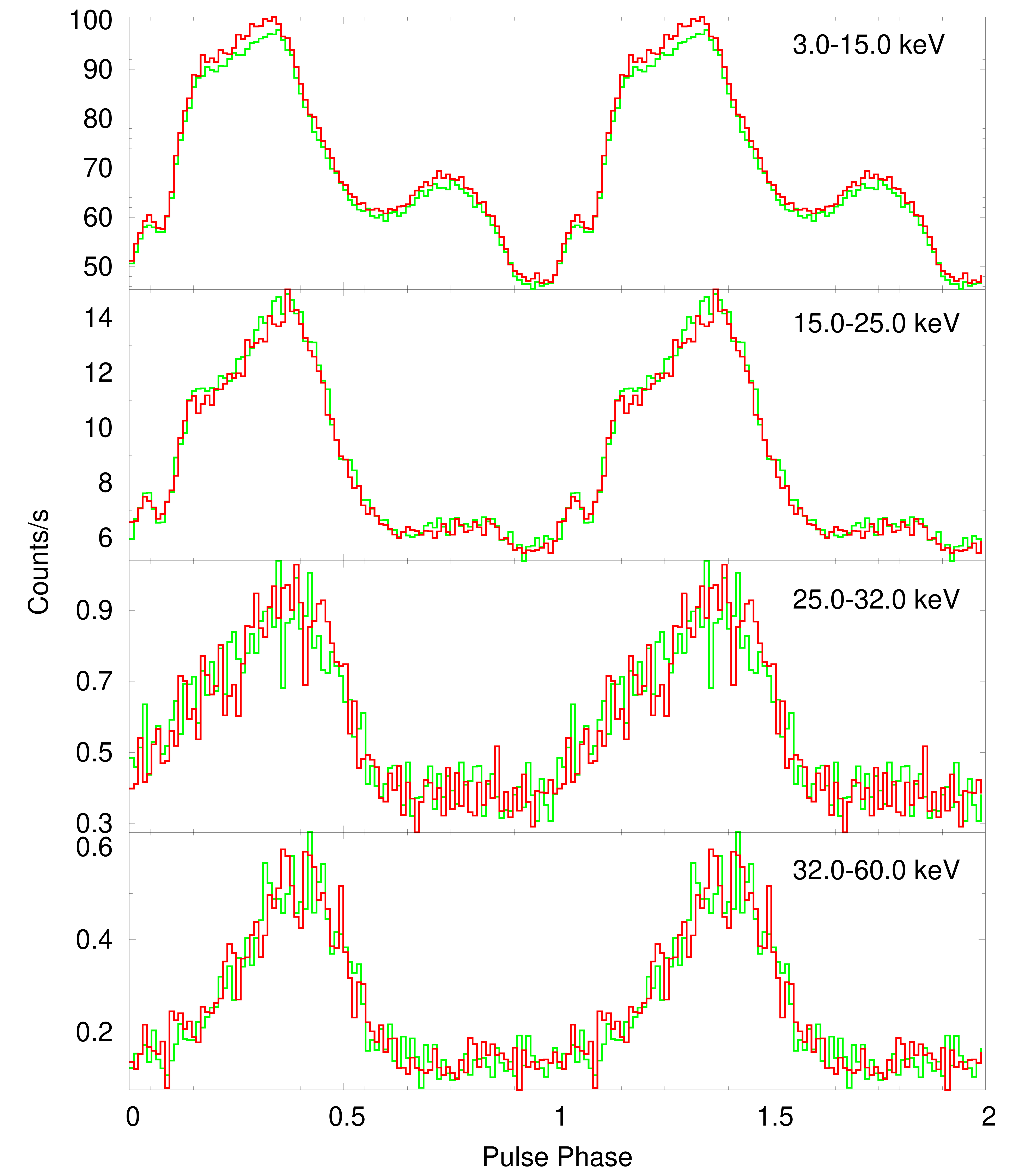}
\caption{Energy resolved pulse profiles for \src~ from \nustar~ FPMA (green) and FPMB (red).}
\label{pp_energy_nu}
\end{figure}

\subsubsection{\nustar~}
For timing analysis with \nustar~data, 20 ms binned \nustar~ lightcurves were extracted from the barycentric corrected event files using the task \texttt{nuproducts}. Similar to \suzaku~lightcurves, the light travel time due to orbital motion were corrected for each \nustar~lightcurves by using the orbital parameters from \citet{falagna2015} extrapolated to the time of the \nustar~observation. 
During this observation, the pulse period was determined using the epoch folding technique ( with the tool \texttt{efsearch} in \texttt{FTOOLS} ). The period derivative was found to be consistent with zero.
\\
To investigate the energy dependence of pulse profile, energy resolved pulse profiles were generated by folding the light curves over the best period determined (4.8026 $\pm 0.0002$ s) in four different energy bands: 3.0-15.0 keV, 15.0-25.0 keV, 25.0-32.0 keV and 32.0-60.0 keV. The pulse phase was chosen to have the phase of the pulse peak aligned with the same in the \suzaku~ pulse profile. Fig.~\ref{pp_energy_nu} shows the resulting pulse profiles. At $\phi \sim$ 0.75, a secondary pulse peak is visible in the low energy range (3.0-15.0 keV) which decreases as we go to higher energy ranges. An additional narrow and small peak is also visible at $\phi \sim$ 0.05 below 25 keV similar to the feature observed in \suzaku~observation. While the pulse profile has these two additional features in the low energy bands, it is seen that the pulse profile in the higher energy bands has single-peaked shape. No variation in the position of the primary peak was observed.

\subsection{Spectral analysis}
\label{spectral analysis}

\subsubsection{\suzaku~}
\begin{figure*}
\centering
\includegraphics[width=\textwidth]{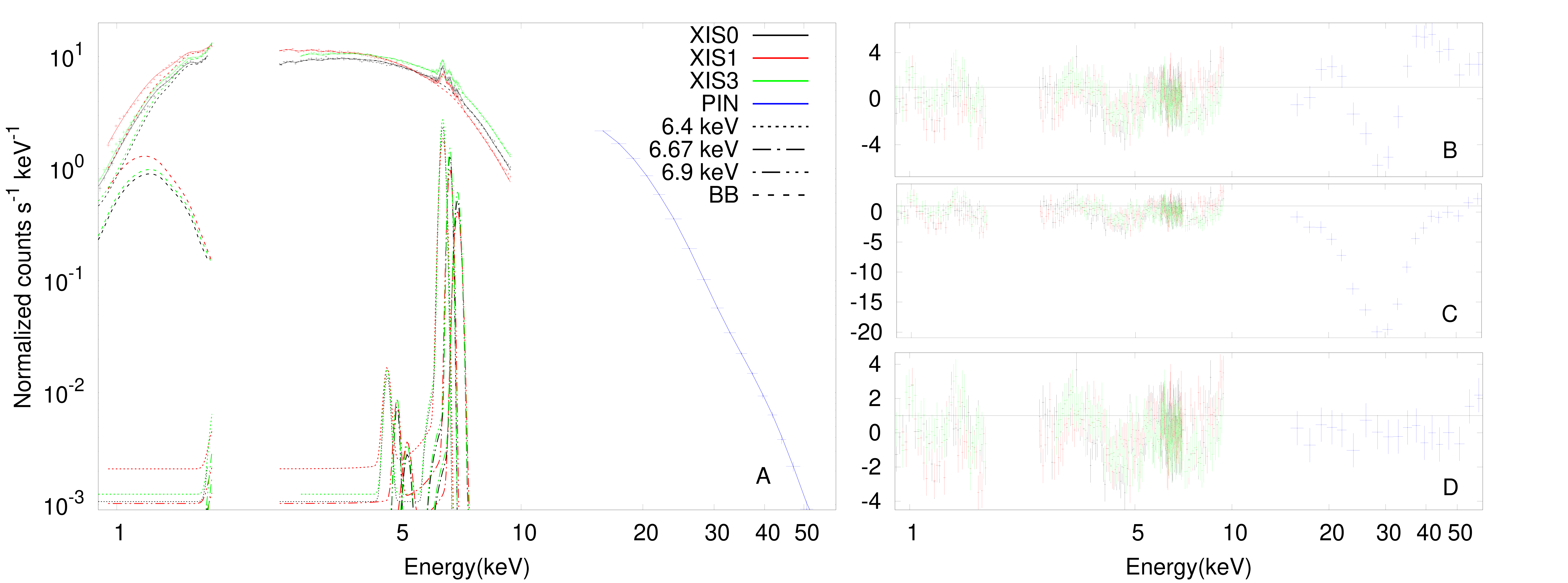}
\caption{ Panel A shows the broadband continuum in 0.8-60.0 keV energy range obtained from \suzaku~with the best-fit model \texttt{powerlaw*newhcut} with a Gaussian absorption feature (\texttt{gabs} for cyclotron line), iron emission lines, and an additional blackbody component. Panel B corresponds to the residual to the model fit without \texttt{gabs}. Panel C shows residuals to the model setting strength of cyclotron line to zero, but leaving the other parameters at their best fit values and panel D corresponds to the residuals to the best fit model with \texttt{gabs}. The residuals are plotted in unit of the error bar for each bin, representing their contribution to the chi-squared.}
\label{spec_su}
\end{figure*}

We performed pulse phase averaged spectral analysis of \src~using data from three XIS (0, 1 and 3) and the HXD/PIN (shown in Figure \ref{spec_su}). Spectral fitting was performed using \texttt{XSPEC} v12.10.0e. Artificial structures are known in the XIS spectra around the Si
edge and Au edge and the energy range of 1.75-2.23 keV is usually not used for spectral fitting. We fitted the spectra simultaneously with all parameters tied, except the relative instrument normalizations which were kept free. The 2048 channel XIS spectra were rebinned by a factor of 6 upto 4.2 \rmfamily{keV}, by 2 from 4.2-4.7 \rmfamily{keV} by 14 for the rest. The PIN spectra were binned by a factor of 4 till 22.5 \rmfamily{keV}, by 6 from 22.5-45 \rmfamily{keV}, and by 10 for the rest. 

The hard X-ray spectral continuum is typically described by different phenomenological models consisting of a power-law with high-energy cutoffs of various functional forms ('highecut', 'fdcut' and 'newhcut') usually adopted for highly magnetised accreting neutron stars. Other models based on the work by \cite{Becker_Wolff} that directly models the physical processes in the accretion column leading to the production of hard X-ray continuum are being tested and revised. We fit the X-ray spectrum of \src~with such standard continuum models \footnote{http://heasarc.gsfc.nasa.gov/xanadu/xspec/manual/XspecModels.html} used for NS-HMXBs ~like \texttt{CUTOFFPL}, \texttt{COMPTT} and \texttt{NEWHCUT}(\citealt{Burderi}). All these models provided comparable fits with the lowest reduced $\chi^{2}$ given by a smooth high energy cutoff model (\texttt{NEWHCUT}). We therefore use this model for further fitting. To the continuum model, we also add a blackbody component to take care of the soft excess below 2\, keV, a feature ubiquitous to HMXBs \citep{paul02, hickox2004} and also used by various authors to specifically fit the Cen X-3 spectrum \citep{Burderi, Naik_2011}. In addition to this, the XIS spectrum of Cen X-3 is also rich in Fe lines caused by fluorescent emission of primary X-rays from the neutron star with the surrounding matter \citep{Naik_2011}. We include three Gaussian in the fits to account for the iron emission lines at 6.4 keV, 6.7 keV, and 6.97 keV. Finally, to describe the CRSF, which is clearly seen in the X-ray spectrum, we used a multiplicative absorption line with a Gaussian optical depth profile (\texttt{gabs} in \texttt{XSPEC}). The centroid energy of CRSF for the best-fit is ${E_{cyc} = 30.29^{+0.68}_{-0.61}}$ keV. A complete list of spectral parameters used for this fitting is shown in Table \ref{suzaku_spectrum} and the 0.8-60 keV \suzaku~spectrum is shown in Figure \ref{spec_su}. The 2-10 keV flux for the best-fit model with \texttt{gabs} is $3.18 \times 10^{-9}$ ergs cm$^{-2}$ s$^{-1}$.

\begin{table}
\centering
\begin{threeparttable}[t]
  \caption{Suzaku: Results of the fit in the energy range 0.8-60.0 keV}
  \label{suzaku_spectrum}
       \begin{tabular}{|l|l|l|}
    \hline
    \hline
    Parameter & Value\tnote{1} & Value\tnote{2} \bigstrut\\
    \hline
    \hline
    nH & $1.53\pm 0.02  \times 10^{22} cm^{-2}$ & $1.53\pm 0.02\times 10^{22} cm^{-2}$ \bigstrut\\
    
    kT & $0.118\pm 0.003$ keV & $0.118\pm 0.003$  keV\bigstrut\\
    
    Norm$_{bb}$\tnote{3} & $0.031^{+0.004}_{-0.003}$ & $0.030\pm 0.004$\bigstrut\\
    
    PhoIndex & $1.037\pm 0.006$ & 1.038$\pm$ 0.006 \bigstrut\\
    
    Norm$_{po}$\tnote{4} & $0.283\pm 0.003$ & $0.284\pm 0.003$ \bigstrut\\
    
    E$_{cut}$ & $15.09^{+0.71}_{-0.79}$ keV & $14.79^{+0.41}_{-0.50}$ keV  \bigstrut\\
    
    E$_{fold}$ & $8.33^{+0.20}_{-0.19}$  keV & $7.22\pm 0.09$  keV\bigstrut\\
    
    Smoothing width\tnote{5} & 5.0 keV & 5.0 keV\bigstrut\\
    
    E$_{cyc}$ & $30.29^{+0.68}_{-0.61}$ keV & - \bigstrut\\
    
    $\sigma_{cyc}$ &  $5.00^{+0.96}_{-0.79}$ keV & - \bigstrut\\
    
  
    $\tau_{cyc}$ & 0.396 $\pm$ 0.128 & - \bigstrut\\
    
    E$_{Fe}$ & $6.402\pm 0.003$ keV & $6.402^{+0.002}_{-0.003}$ keV\bigstrut\\
    
    $\sigma_{Fe}$ & 0.047  keV & 0.047  keV\bigstrut\\
    
    Eqw$_{Fe}$ & 0.085$^{+0.03}_{-0.02}$ keV & 0.0852$^{+0.02}_{-0.03}$ keV\bigstrut\\
    
    E$_{Fe}$ & 6.67 keV & 6.67 keV\bigstrut\\
    
    $\sigma_{Fe}$ & 0.006  keV & 0.006  keV\bigstrut\\
    
    Eqw$_{Fe}$ & 0.031$^{+0.003}_{-0.004}$ keV & 0.031$\pm 0.03$ keV\bigstrut\\
    
    E$_{Fe}$ & 6.97 keV & 6.97 keV\bigstrut\\
    
    $\sigma_{Fe}$ & $0.082^{+0.018}_{-0.017}$  keV & $0.083^{+0.018}_{-0.017}$  keV\bigstrut\\
    
    Eqw$_{Fe}$ & 0.041$\pm 0.002$  keV & 0.041$\pm 0.02$ keV\bigstrut\\
    
    Flux\tnote{6} & $3.18 \times 10^{-9}$ & $3.18 \times 10^{-9}$\bigstrut\\   
    
    \hline
    ${\chi_{red}^{2}}/${d.o.f} & 1.63/635 & 1.99/638 \bigstrut\\
    \hline
    \end{tabular}
    \begin{tablenotes}
    \item[Note: The error bars are quoted at 90\% confidence limit.] 
    \item[1] Model: \texttt{phabs*(powerlaw*newhcut*gabs + bb + ga + ga + ga)}
    \item[2] Model: \texttt{phabs*(powerlaw*newhcut + bb + ga + ga + ga)}
    \item[3] in units of L$_{39}$/D$_{10}^{2}$ where L$_{39}$ is source luminosity in units of 10$^{39}$ ergs/s and D$_{10}$ is source distance in units of 10 kpc.
    \item[4] in units of photons/keV/cm$^{2}$/s at 1 keV.
    \item[5] in the energy range of E$_{cut}\pm5$ keV, a third order polynomial function is used to have a smooth high energy cut-off.(\citealt{Burderi})
    \item[6] 2-10 keV flux in units of ergs cm$^{-2}$ s$^{-1}$. 
    \end{tablenotes}
    \end{threeparttable}
\end{table}

\subsubsection{\nustar}

\begin{figure*}
\centering
\includegraphics[width=\textwidth]{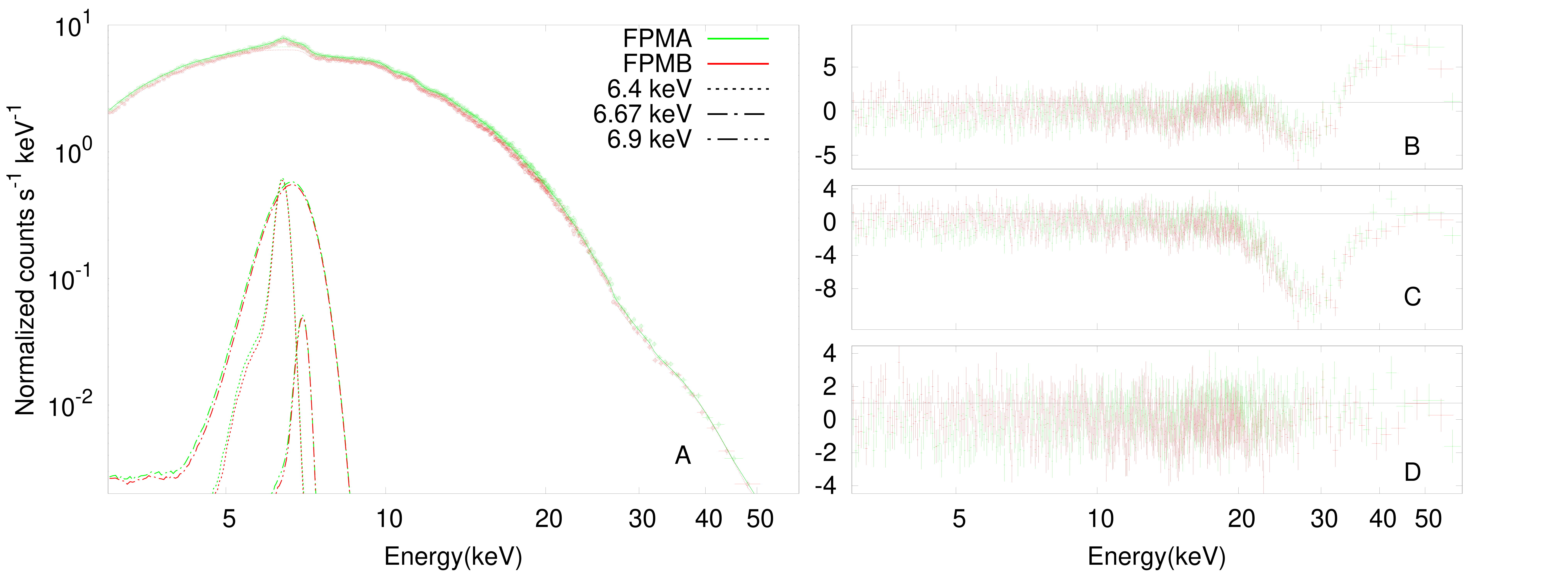}
\caption{Panel A shows the broadband continuum in in 3.0-60.0 keV energy range obtained from \nustar~with the best-fit model \texttt{powerlaw*newhcut*pcfabs} with a Gaussian absorption feature (\texttt{gabs} for cyclotron line) and iron emission lines. Panel B corresponds to the residuals to the model fit without gabs. Panel C shows residuals to the model setting strength of cyclotron line to zero, but leaving the other parameters at their best fit values and panel D corresponds to the residuals to the best fit model with \texttt{gabs}.
The residuals are plotted in unit of the error bar for each bin, representing their contribution to the chi-squared.}
\label{spec_nu}
\end{figure*}

We carried out spectral fitting of the pulse phase averaged \nustar ~ FPMA and FPMB spectra simultaneously, leaving a normalization constant free to vary between the two spectra. The spectra were fitted in the energy range 3.0-60.0 keV. The energy ranges below and above this were neglected due to limited statistics. We grouped the spectral data from both FPMA and FPMB to collect at least 30 counts per bin. The best spectral fitting was obtained with \texttt{NEWHCUT} (\citealt{Burderi}). The photoelectric absorption in the low energy part of the spectra was taken into account using the \texttt{phabs} model in \texttt{XSPEC}. Due to no spectral coverage of \nustar~ below 3 keV, the value of nH was frozen at the Galactic hydrogen column density towards the source, 1.1 $\times 10^{22}$ cm $^{-2}$ \footnote{https://heasarc.gsfc.nasa.gov/cgi-bin/Tools/w3nh/w3nh.pl}. A blackbody component with a temperature of $\sim$ 0.118 keV can also not be constrained. We therefore added a partial covering absorption model (\texttt{pcfabs} in \texttt{XSPEC}) to take local absorption into account. Since the resolution of \nustar~ is not high enough to clearly resolve the three iron emission lines at 6.4 keV, 6.67 keV and 6.97 keV (\citealt{Ebiwasa}), we modelled these as additive Gaussian lines \citep{Naik_2011} with fixed centroid line energies allowing the respective sigma and normalizations to vary. The equivalent width of each is determined. The emission line at 6.97 keV is not significant in the \nustar~spectrum likely due to low resolution of the instrument. The residuals in panel B (see Figure \ref{spec_nu}) suggests a cyclotron resonant scattering feature. To investigate this, we used a multiplicative absorption line with a Gaussian optical depth profile (\texttt{gabs} in \texttt{XSPEC}). The centroid energy of CRSF for the best-fit is $E_{cyc}$ = ${29.22^{+0.28}_{-0.27}}$ keV. The addition of \texttt{gabs} reduced the ${\chi_{red}^{2}}/{d.o.f}$ of the fit from 1.714/1633 to 1.097/1630 (Table \ref{par_nu}) . The 2-10 keV flux for the best-fit model with \texttt{gabs} is  ${1.589 \times 10^{-9}}$ ergs cm$^{-2}$ s$^{-1}$.

\begin{table}
\centering
\begin{threeparttable}[t]
   
    \caption{NuSTAR:Results of the fit in the energy range 3.0-60.0 keV}
    \label{par_nu}
    \begin{tabular}{|l|l|l|}
    \hline
    \hline
      Parameter & Value\tnote{1} & Value\tnote{2} \bigstrut\\
    \hline
    \hline
    NH1 & $1.10 \times 10^{22} $cm$^{-2}$ & $1.1 \times 10^{22}$ cm$^{-2}$ \bigstrut\\
    
    NH2  & 21.65$^{+1.07}_{-1.12} \times 10^{22} $cm$^{-2}$ & 22.10$^{+1.05}_{-1.07} \times 10^{22} $cm$^{-2}$ \bigstrut\\
    
    CvrFrac & 0.590$\pm 0.009$ & 0.601$\pm 0.009$ \bigstrut\\
     
    PhoIndex & $1.21\pm 0.01$ & $1.23\pm 0.01$ \bigstrut\\
    
    Norm$_{po}$\tnote{3} & $0.275\pm 0.013$ & $0.289^{+0.014}_{-0.013}$ \bigstrut\\
    
    E$_{cut}$ & $14.14\pm 0.06$ keV & $14.53\pm 0.05$ keV  \bigstrut\\
    
    E$_{fold}$ & $8.88\pm 0.10$  keV & $7.56\pm 0.05$  keV \bigstrut\\
    
    Smoothing width\tnote{4} & 5.0 keV & 5.0 keV \bigstrut\\
    E$_{cyc}$ & $29.22^{+0.28}_{-0.27}$ keV & - \bigstrut\\
    $\sigma_{cyc}$ &  $4.37^{+0.25}_{-0.24}$ keV& - \bigstrut\\
    
    $\tau_{cyc}$ & 0.417 $\pm$ 0.044 & - \bigstrut\\
    
    E$_{Fe}$ & 6.4 keV & 6.4 keV \bigstrut\\
    
    $\sigma_{Fe}$ & 0.007  keV & 0.007  keV \bigstrut\\
    
    Eqw$_{Fe}$ & 0.038$\pm 0.004$  keV & 0.037$\pm 0.004$ keV \bigstrut\\
    
    E$_{Fe}$ & 6.67 keV & 6.67 keV \bigstrut\\
    
    $\sigma_{Fe}$ & 0.53$\pm 0.03$  keV & 0.53$^{+0.04}_{-0.03}$  keV \bigstrut\\
    
    Eqw$_{Fe}$ & 0.117$^{+0.011}_{-0.007}$  keV & 0.113$^{+0.009}_{-0.009}$ keV \bigstrut\\
    
    E$_{Fe}$ & 6.97 keV & 6.97 keV \bigstrut\\
    
    $\sigma_{Fe}$ & 0.0003 keV & 0.0003 keV \bigstrut\\
    
    Eqw$_{Fe}$ & 0.00326$\pm 0.00001$keV & 0.00326$^{+0.00001}_{-0.00002}$ keV \bigstrut\\
    
    Flux\tnote{5} & $1.589 \times 10^{-9}$ & $1.589 \times 10^{-9}$ \bigstrut\\
    \hline
  ${\chi_{red}^{2}}/${d.o.f} & 1.097/1630 & 1.714/1633\bigstrut \\
    \hline

    \end{tabular}
    \begin{tablenotes}
    \item[Note: The error bars are quoted at 90\% confidence limit.] 
    \item[1] Model: \texttt{phabs*(powerlaw*newhcut*gabs*pcfabs + ga + ga + ga)}
    \item[2] Model: \texttt{phabs*(powerlaw*newhcut*pcfabs + ga + ga + ga)}
    \item[3] in units of photons/keV/cm$^{2}$/s at 1 keV.
    \item[4] in the energy range of E$_{cut}\pm5$ keV, a third order polynomial function is used to have a smooth high energy cut-off.(\citealt{Burderi})
    \item[5] 2-10 keV flux in units of ergs cm$^{-2}$ s$^{-1}$. 
    \end{tablenotes}
    \end{threeparttable}
\end{table}

\section{Discussion}
\label{discussion}
\begin{figure}
\centering
\includegraphics[width=0.5\textwidth]{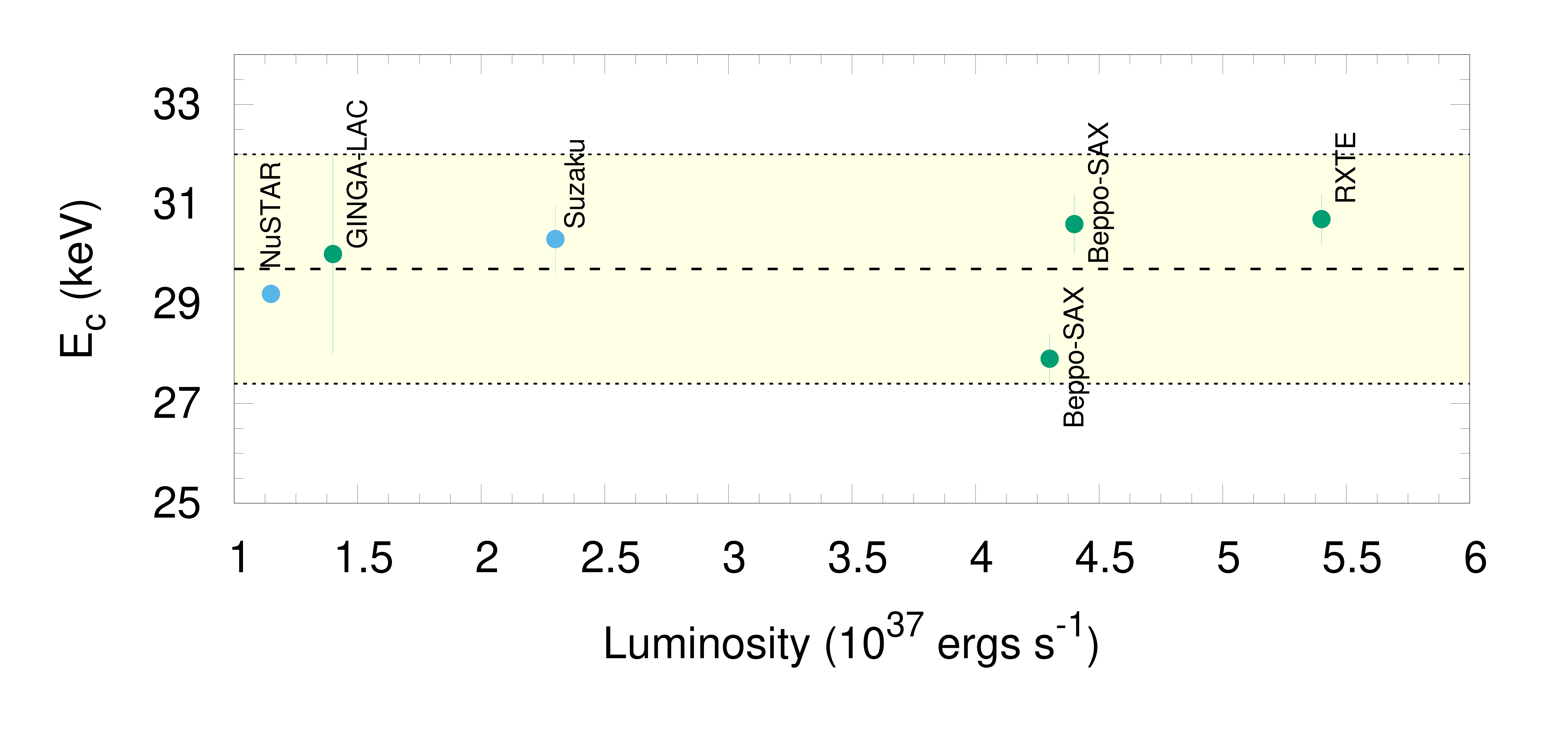}
\caption{E$_c$ vs Luminosity in the energy band 2.0-10.0 keV (See table \ref{cycl_line} for details). The dashed line here represents the average value while the dotted lines correspond to the maximum CRSF energy variation around the average value.}
\label{lx_cyc}
\end{figure}

We have presented results from a detailed timing and spectral analysis of the HMXB Cen X-3 with \suzaku~and \nustar. The pulse profiles are found to vary with energy. Although similar dependence was presented in previous works by \citet{Suchy_2008} and \citet{raichur2010}, we observed an additional feature, a bump at $\phi \sim$ 0.05 below 25 keV in both \suzaku~and \nustar~observations. A similar bump at $\phi \sim$ 0.05 was observed in 0.1-1.8 keV energy range by \citet{Burderi}.
\\
We also presented measurements of the cyclotron line feature in the \suzaku~and \nustar~data. In both \suzaku~and \nustar~spectra, we use the gaussian absorption model \texttt{GABS} to find the cyclotron line at 30.29 keV and 29.22 keV respectively. 
\\
To look for any variation of cyclotron line with luminosity, we used the line measurements from earlier observations of \src~. The energy of cyclotron resonance scattering feature E$_{c}$ in the pulse averaged X-ray spectrum of Cen X-3 as reported in several papers since its discovery is presented in Table \ref{cycl_line}. The flux for all the previous measurements were reported for 2-10 keV energy band except for \ginga~. For the comparison, the 2-30 keV energy band flux reported for \ginga~ measurement was corrected in 2.0-10.0 keV energy range using 
\ensuremath{f_c = f^{po} \times \frac{f_{R}}{f^{po}_{R}}} where $f^{po}$ and $f^{po}_{R}$ are the power law model flux in 2-10 keV band and 2-30 keV energy range respectively with $\Gamma$=1.01 and norm$_{po}=1$ and $f_R$ is the flux in the 2-30 keV energy range as reported in the paper.
It is noted that different continuum and cyclotron line models were used to fit the spectra from different instruments. To allow a comparison of the values obtained with the previous measurements, we fitted the \nustar~spectrum with the same models as adopted in these works for the spectrum obtained with \ginga, ~\rxte~ and \beppo~(see Table \ref{cycl_line}). We obtained a CRSF energy value of 29.2 \, keV with the models used for \rxte~ and \beppo~ spectrum which is consistent with the value obtained with our \nustar~ model. For the \citet{Nagase} model used for \ginga~ spectrum, in the final fit, the high-energy cutoff was replaced by Lorentzian cyclotron line component (\texttt{cylabs}; \citet{cyc1}, \citet{cyc2}). This model provided a value of 27.5 \, keV for \nustar~ spectrum compared to the value 29.2 \, keV obtained with \texttt{gabs}. The CRSF energy measured with \ginga~ has larger error and it is also less certain as the spectrum was just upto above 30 \, keV and barely covered the cyclotron line. We also note here that the \ginga~ results showed a much larger optical depth of the CRSF compared to all the other observations. As the spectrum did not cover as high energy as the other instruments like \rxte~, \beppo~, \nustar~ and \suzaku~, it also did not require a high energy cutoff.
We notice that except one CRSF measurement with Beppo-SAX (\citealt{Santangelo}), all are consistent with a CRSF energy in the range of  29.5-30.0 keV over a luminosity range of 1.1-5.4$\times10^{37}$ergs s$^{-1}$,  while sources like V 0332+52 shows about 20$\%$ CRSF energy variation in the same luminosity range  (\citealt{Tsy}). A lower CRSF value in one of the \beppo~ observations, however, corresponds to a luminosity that is in the middle of this entire luminosity range and therefore does not indicate a luminosity dependence of the line energy. An average value of E$_{cyc}=$ 29.7 keV was calculated considering the minimum and the maximum observed cyclotron line energies with the respective uncertainties. Including all the reported results, the upper limit on CRSF energy variation between measurements is $\pm$ 7.7\% ($\sim$ 2.3 keV) around the average value including the statistical errors and any systematic errors in each of the measurements. We do not find a correlation of the cyclotron line energy with luminosity (Fig.~ \ref{lx_cyc}) for Cen X-3. While comparing spectral parameters like the CRSF using data from different observations, inter-calibration of different instruments also needs to be considered. In Cen X-3, the variation between different measurements is within 2.3 keV of the average value while in 4U 1538-52, the variation is within about 1.5 keV of the 
average value (\citealt{varun}, \citealt{hemphill}). Lack of detection of significant variation in the cyclotron line energy also indicates good inter-calibration of the different instruments within a few percent.
\\Such studies for the variation of cyclotron line with luminosity has been carried out earlier for other X-ray pulsars, notable ones being Her X-1 and Vela X-1 both of which show a positive correlation (see., \citealt{staubert2019} for review) with luminosity. Like in the case of V 0332+53, the energy of the fundamental CRSF in another brightest known CRSF source, SMC X-2 (\citealt{Jaiswal}) is negatively correlated with L$_X$, while Her X-1 (\citealt{staubert2007}), GX 304-1 (\citealt{Klochkov}; \citealt{Malacaria};\citealt{Rothschild_16}), A 0535+26 \citep{sartore_15}, Vela X-1 \citep{furst_!4}, Cep X-4 \citep{vyornov_18} show a positive correlation. There are also X-ray pulsars with tentative evidence of correlation between CRSF energy and luminosity but so far not confirmed, such as 4U 0115+63 \citep{iyer_15} and 4U 1538-522 \citep{hemphill_14}. It is now a general consensus that the variation of cyclotron line energy is due to the change in the line formation region above the neutron star surface with X-ray luminosity \citep{becker2012}. For typical luminosities less than 10$^{37}$ erg s$^{-1}$, the matter stops close to the stellar surface and therefore, the location of line
 forming region is independent of X-ray flux. Above this luminosity, matter is stopped well above the stellar surface because of shock formation (Coulomb interaction dominated). Since the ram pressure increase with increased accretion rate, it pushes the shock region towards the neutron star surface where the magnetic field is higher.
 This causes the positive correlation of line energy with flux \citep{staubert2007}. Furthermore, if the Eddington luminosity exceed the critical luminosity ( $\sim$ 3 $\times$ 10$^{37}$ erg s$^{-1}$ ) for typical neutron star parameters, a radiation-dominated shock forms. With the increase in accretion rate, this shock formation region move away from the stellar surface and an anti-correlation of the CRSF energy with luminosity is expected. 
 \\
 A long-term decay in CRSF energy between 1992 to 2012 was assessed in Her X-1 (\citealt{staubert_2014}). Similar to Her X-1, long-term decrease was found in Vela X-1 (\citealt{la_2016}). The observed decrease could be attributed to the change of magnetic field configuration at the polar cap or the geometric displacement of the line forming region (\citealt{staubert_2014}). On the contrary, a positive trend with 5$\%$ increase in CRSF energy from 1996 to 2012 was observed in 4U 1538-52 (\citealt{hemp_2016}).
Although the evidence of a long-term change in CRSF energy has been seen in few sources until now, previous works have noted that the centroid line energy in Cen X-3 is stable over the past 14 years, at 31.6$\pm$ 0.2 keV as seen in the \swift/BAT data \citep{Ji2019}. The current work examines the possible variation of the cyclotron line energy of Cen X-3 over a period of close to 27 years and concludes that there is no evidence for a time variability of the CRSF nor a luminosity dependence.

\begin{table*}
\centering
\begin{threeparttable}
\caption{Cyclotron line observations for Cen X-3}
 \label{cycl_line}
 \centering
     \begin{tabular}{|p{0.5cm}|c|c|c|c|c|}
     
     \hline
     \hline
    \multirow{3}{0.3cm}{S.no.} & \multirow{3}{2.5cm}{Date of observations} & \multirow{3}{2.5cm}{Observatory}  &  \multirow{3}{2.5cm}{Flux\tnote{a}\\(ergs cm$^{-2}$ s$^{-1}$)} & \multirow{3}{2.5cm}{Luminosity\tnote{b}\\(ergs s$^{-1}$)} & \multirow{3}{3cm}{Cyclotron Line Parameters\tnote{c}}\bigstrut

    \\
     & & & &  & 
    \\
    & & & & &   
    \\
    
    \hline
    \hline
    
     \multirow{3}{0.3cm}{1.} & \multirow{3}{2.5cm}{22-24 Mar., 1989} & \multirow{3}{2.5cm}{GINGA-LAC\tnote{1}}  &  \multirow{3}{2.5cm}{$1.9\times 10^{-9}$ \tnote{*}} & \multirow{3}{2.5cm}{$1.4 \times 10 ^{37}$}  &  \multirow{3}{3cm}{$E_c = 30\pm 2$ \\$\sigma = 3.0\pm 0.3$ \\ $\tau=3.0\pm 0.3$ }   \bigstrut
     \\
     & & & & &  
    \\
     & & & & &  
    \\ 
    
    \hline
    
    \multirow{4}{0.3cm}{2.} & \multirow{4}{2.5cm}{17 Feb., 1997} & \multirow{4}{2.5cm}{BeppoSAX\tnote{2}} &  
    \multirow{4}{2.5cm}{-}  & \multirow{4}{2.6cm}{$4.3 \times 10 ^{37}$} &
    \multirow{4}{3cm}{$E_c= 27.9\pm 0.5$\\ $\sigma = 4.16\pm 0.8$\\$\tau=0.53$}\bigstrut
    \\ 
     & &  & &  &   
     \\
     &  & & &  &   
     \\ 
     & & & & &
     \\
     
    \hline
    
    \multirow{4}{0.3cm}{3.} & \multirow{4}{2.5cm}{27-28 Feb., 1997}& \multirow{4}{2.5cm}{BeppoSAX\tnote{3}} & \multirow{4}{2.5cm}{$5.7\times 10^{-9}$}  &  \multirow{4}{2.6cm}{$4.4 \times 10 ^{37}$} &
    \multirow{4}{3cm}{$E_c= 30.6\pm 0.6$ \\$\sigma = 5.9\pm 0.7$\\ $\tau=0.77$}  \bigstrut
    \\ 
     &  & &  &  &
    \\
     & & &  &  & 
    \\ 
    & & & & &
    \\
    
    \hline
    
    \multirow{4}{0.3cm}{4.} & \multirow{4}{2.5cm}{28 Feb.-3 Mar., 1997} & \multirow{4}{2.5cm}{RXTE-HEXTE (PCA)\tnote{4}} & \multirow{4}{2.5cm}{-} & \multirow{4}{2.6cm}{$5.4 \times 10 ^{37}$} & \multirow{4}{3cm}{$E_c= 30.7^{+0.5}_{-0.4}$\\$\sigma = 6.4{^{+1.0}_{-0.8}}$ \\ $\tau=0.67{^{+0.17}_{-0.07}}$} \bigstrut
    \\ 
    & & &  &  & 
    \\
     & & &  &  & 
     \\
     & & & & & 
     \\
  
    \hline
    
    \multirow{4}{0.3cm}{5.} & \multirow{4}{2.5cm}{8 Dec., 2008}& \multirow{4}{2.5cm}{Suzaku\tnote{5}} & \multirow{4}{2.5cm}{3.18$ \times 10^{-9}$} & \multirow{4}{2.6cm}{2.30 $\times 10^{37}$} & \multirow{4}{3cm}{$E_c=30.29^{+0.68}_{-0.61}$\\$\sigma$ = $5.00^{+0.96}_{-0.79}$\\$\tau$ = 0.396 $\pm$ 0.128}  \bigstrut
    \\ 
     &  & & &  &  
    \\
     & & & &  & 
    \\ 
    & & & & &
    \\
    
    \hline
    
    \multirow{4}{0.3cm}{6.} & \multirow{4}{2.5cm}{30 Nov., 2015}& \multirow{4}{2.5cm}{NuSTAR\tnote{5}} & \multirow{4}{2.5cm}{1.589$ \times 10^{-9}$} & \multirow{4}{2.6cm}{1.149$ \times 10^{37}$} & \multirow{4}{3cm}{$E_c=29.22^{+0.28}_{-0.27}$\\ $\sigma$ = $4.37^{+0.25}_{-0.24}$\\$\tau$ = 0.417 $\pm$ 0.044} \bigstrut
    \\ 
     &  & & &  &    
    \\
     & & & &  &  
    \\ 
     & & & &  &  
    \\ 
    
    \hline
    
    \end{tabular}
 
 \begin{tablenotes}
  \item[*] Corrected flux in 2.0-10.0 keV (See Discussion.)
\\
 \item[a] Flux in the energy range 2.0-10.0 keV.
 \item[b] Assuming a distance of 8 kpc.
 \item[c] $E_c$ and $\sigma$ are in the units of keV. $\tau$ is the optical depth.
 \item[1] Ref: \citet{Nagase}
 \item[2] Ref: \citet{Santangelo}
 \item[3] Ref: \citet{Burderi}
 \item[4] Ref: \citet{Suchy_2008} 
 \item[5] This work.
 \end{tablenotes}
 \end{threeparttable}
 \end{table*}

\section*{Acknowledgements}
We thank the anonymous referee for providing very valuable comments. This research has made use of archival data and software provided by NASA's High Energy Astrophysics Science Archive Research Center (HEASARC)\footnote{\url{https://heasarc.gsfc.nasa.gov/}}, which is a service of the Astrophysics Science Division at NASA/GSFC. The scientific results are based on the data obtained from the \suzaku~satellite, a collaborative mission between the space agencies of Japan (JAXA) and the USA (NASA) and \nustar~mission, a project led by the California Institute of Technology, managed by the Jet Propulsion Laboratory, and funded by the National Aeronautics and Space Administration.

\section*{Data Availability}
The observational data underlying this work is publicly available through High Energy Astrophysics Science Archive Research Center (HEASARC). Any additional information will be shared on reasonable request to the corresponding author.



\bibliographystyle{mnras}
\bibliography{reference_cenx3} 


\bsp	
\label{lastpage}
\end{document}